\documentclass[twocolumn,english,prb,superscriptaddress,showpacs]{revtex4}
\usepackage{graphicx}

\newcommand{\CMS}{CoMnSi}
\newcommand{\CNMS}{Co$_{0.95}$Ni$_{0.05}$MnSi}
\newcommand{\CMFS}{CoMn$_{0.95}$Fe$_{0.05}$Si}
\newcommand{\CMCS}{CoMn$_{0.98}$Cr$_{0.02}$Si}

\newcommand{\TN}{$T_{N}$}
\newcommand{\done}{$d_{\rm{1}}$}
\newcommand{\dtwo}{$d_{\rm{2}}$}

\begin{document}

\title{Magneto-elastic coupling and competing entropy changes in substituted CoMnSi metamagnets}
\author{A. Barcza}
\affiliation{Dept. of Materials Science and Metallurgy, University of Cambridge, New Museums Site, Pembroke Street, Cambridge, CB2 3QZ, UK}
\author{Z. Gercsi}
\affiliation{Dept. of Physics, Blackett Laboratory, Imperial College London, London SW7 2AZ UK}
\author{H. Michor}
\affiliation{Institute of Solid State Physics, Vienna University of Technology, A-1040, Wien, Austria}
\author{K. Suzuki}
\affiliation{Dept. of Materials Engineering, Monash University, Clayton, VIC 3800, Australia }
\author{W. Kockelmann}
\affiliation{ISIS, Rutherford Appleton Laboratory, Oxon OX11 0QX, UK}
\author{K.S. Knight}
\affiliation{ISIS, Rutherford Appleton Laboratory, Oxon OX11 0QX, UK}
\author{K.G. Sandeman}
\affiliation{Dept. of Physics, Blackett Laboratory, Imperial College London, London SW7 2AZ UK}

\begin{abstract}
We use neutron diffraction, magnetometry and low temperature heat capacity to probe giant magneto-elastic coupling in CoMnSi-based antiferromagnets and to establish the origin of the entropy change that occurs at the metamagnetic transition in such compounds.  We find a large difference between the electronic density of states of the antiferromagnetic and high magnetisation states.  The magnetic field-induced entropy change is composed of this contribution and a significant counteracting lattice component, deduced from the presence of negative magnetostriction.  In calculating the electronic entropy change, we note the importance of using an accurate model of the electronic density of states, which here varies rapidly close to the Fermi energy.
\end{abstract}
\maketitle
Magnetic field-driven phase transitions have long been of interest in studies of magnetoresistance~\cite{ramirez_1997a} and magnetic shape memory.~\cite{vasiliev_1999a}  Recently, research into the magnetocaloric effect (MCE) in the vicinity of such transitions has been revived, fuelled by interest in solid state, gas-free methods of cooling.~\cite{sandeman_2011a,turley_2012a}  Most works that examine materials with large MCEs focus on Curie transitions in ferromagnets such as Gd$_{5}$(Si,Ge)$_{4}$~\cite{pecharsky_1997a}, (Mn,Fe)$_{2}$(P,Z)~\cite{tegus_2002a} and La(Fe,Si)$_{13}$.~\cite{fujita_2003a}  They exhibit a conventional MCE, namely a positive change of temperature when an increasing magnetic field is applied. However, a smaller set of inverse magnetocaloric materials have also attracted interest.  In these, the MCE is associated with a field-induced transition to a high magnetisation state that exists above, rather than below, a critical temperature.  As a result the magnetic field causes a decrease in temperature.  Examples include Heuslers in which re-entrant ferromagnetism appears with increasing temperature due to the the presence of a structural transformation to a phase with elevated Curie temperature~\cite{krenke_2005a, krenke_2007a} and metamagnetic antiferromagnets such as Mn$_{3}$GaC~\cite{tohei_2003a}, Fe-Rh~\cite{annaorazov_1992a} and CoMnSi~\cite{sandeman_2006a}, the subject of this article.  

Whether the MCE is conventional or inverse, its room temperature magnitude is normally enhanced when there is significant magneto-elastic coupling.  This coupling brings about a first order magnetic transition that releases a significant fraction of the available entropy over a narrow temperature window.    Of the inverse magnetocalorics, Fe$_{0.49}$Rh$_{0.51}$ still holds the record for the magnitude of the adiabatic temperature change per Tesla of applied field~\cite{annaorazov_1992a}, due to its close-to-optimal value of $\partial T_{t} / \partial H$, the rate at which the metamagnetic transition temperature changes with applied field.~\cite{sandeman_2012a}  An investigation of magneto-elastic coupling in inverse magnetocaloric materials is therefore expected to shed light on mechanisms for achieving large MCEs, partly through tuning $\partial T_{t} / \partial H$ to more optimal values than are seen in most ferromagnetic materials.

Recently we used high resolution neutron scattering to observe giant magneto-elastic coupling in CoMnSi and {\CNMS}, associated with the temperature evolution of the antiferromagnetic state of both materials.  In each, the nearest neighbour Mn-Mn separations were shown to change by about 2\% over a 150~K range, the largest such change seen in a metallic magnet.~\cite{barcza_2010a}   Furthermore, application of a magnetic field suppresses the helical antiferromagnetism and yields a metamagnetic transition to a high magnetisation state.  For temperatures close to the N\'{e}el temperature, this transition is continuous.  At lower temperatures the metamagnetic transition couples with the underlying change in Mn-Mn separation and goes through a tricritical point to become first order, accompanied by an enhanced inverse MCE.  Such tricriticality is a useful property, combining large entropy changes with low hysteresis.  Indeed all of the first order magnetic refrigerants (La(Fe,Si)$_{13}$-based, (Mn,Fe)$_{2}$P-based, and manganites) currently trialled in prototype magnetic cooling engines exhibit field-induced critical points.

Since metamagnetism is known to appear with doping as an intermediate state between ferromagnetism and antiferromagnetism in other {\it Pnma} Mn-based alloys, our findings in CoMnSi led us to a simple theoretical model for the existence of either ferromagnetic or antiferromagnetic groundstates in such Mn-based structures, based on how close their nearest neighbour Mn-Mn separations are to a critical value of around 3~\AA.~\cite{gercsi_2010a}  That theoretical work has also been widened to the prediction and synthesis of new CoMn(P,Ge) metamagnets from ferromagnetic end-members, by tuning the nearest neighbour Mn-Mn distance towards the same critical value.~\cite{gercsi_2011a}

In this article we extend our analysis of giant magneto-elastic coupling in CoMnSi to a series of substituted materials in which local exchange interactions, and hence metamagnetism, are altered.  We demonstrate the generic presence of giant magneto-elastic coupling in all compounds studied, and how its magnitude affects the sensitivity of the antiferromagnetic state to an applied magnetic field, and hence tricriticality.  Furthermore, we examine the nature of the metamagnetic transition and its entropic constitution by analysing heat capacity and magnetostriction data.  We show that there is a large change of electronic entropy at the transition that is partially compensated by a smaller change in lattice entropy and that the entropic balance corroborates our density functional model of the V-shaped bandstructure of CoMnSi alloys near to the Fermi energy.

The remainder of this article is organised as follows: section~\ref{sec:Experimental} gives details of our synthesis and characterisation work.  Results are given in Section~\ref{sec:Results} and discussed in Section~\ref{sec:Discussion}.

\section{Experimental}
\label{sec:Experimental}
The samples that we study here were formed by co-melting appropriate amounts of high purity elements Co (99.95 \%), Mn (99.99 \%), Ni (99.994 \%), Fe (99.995 \%), Cr (99.995 \%) and Si (99.9999 \%) under argon in an induction furnace, followed by post-annealing and slow cooling.  Details are given elsewhere.~\cite{sandeman_2006a}  An oxide layer on the as-received manganese was removed by chemical etching.~\cite{fenstad_2000a}  In order to avoid potential embrittlement by a martensitic transition between the orthorhombic groundstate and a high temperature hexagonal state, the alloy was cooled slowly during solidification.  The temperature of the martensitic transition was established by simultaneous differential thermal analysis (SDT) in a TA Instruments Q600.  After each alloy ingot was first formed it was wrapped in tantalum foil, vacuum sealed in a silica tube, and annealed for 60 hours at either 1123~K or 1223~K to fully relieve strain, before cooling to room temperature at 0.2~Kmin$^{-1}$.   

Magnetic measurements in fields of up to 9~Tesla were performed on polycrystalline samples with a Cryogenic Ltd. vibrating sample magnetometer (VSM) and a Quantum Design PPMS VSM.  AC Susceptibility measurements in a Lake Shore 7130 susceptometer were performed over a range of frequencies and between 20 and 320~K.  The N\'{e}el transition temperature of several alloys was measured in zero magnetic field in a conventional heat flux differential scanning calorimeter (DSC, TA Instruments Q2000) with a temperature range of 120 K to 873 K. Small amounts of alloy ($\sim$5 mg) were encapsulated in an aluminium pan and an empty aluminium pan was used as a reference sample. The heating rate in the DSC was set to 10~Kmin$^{-1}$ and data was recorded between 273~K and 673~K.  Heat capacity measurements were also employed in a separate apparatus to determine the Debye temperature and the Sommerfeld coefficient. That apparatus was designed for experiments at low temperatures under quasi-adiabatic conditions.  A detailed description of the experimental setup is given elsewhere.~\cite{manalo_2001a} The calorimeter consisted of an adiabatically shielded sample holder equipped with a resistive heater (a strain gauge) and a thermometer (a Cernox sensor).  Quasi-adiabatic conditions, and hence accurate operation of the calorimeter were limited to temperatures between 1.5 K and 180 K.

Structural characterisation was carried out by X-ray and neutron diffraction.  We conducted a Rietveld refinement of data from room temperature X-ray diffraction using Cu-K$\alpha$ radiation.  X-ray diffraction at room temperature showed that most samples were single phase to within experimental resolution apart from those indicated accordingly in Table~\ref{StructureTable}.   As in our previous work~\cite{sandeman_2006a}, structural refinements of these data agreed well with neutron diffraction data results close to room temperature.  We therefore only present unit cell refinements obtained by high resolution neutron diffraction.

Neutron diffraction was carried out at the time-of-flight High Resolution Powder Diffractometer (HRPD) and at the General Materials Diffractometer (GEM) at ISIS, UK. The former has a resolution of  $\Delta d/d \sim 1\times10^{-4}$ and was used at temperatures between 4.2~K and 500~K.  The latter is capable of hosting a 7~Tesla magnetic field and was used for magnetostriction studies of a CoMnSi ingot between 150~K and 300~K.  The choice of a single ingot was to limit sample movement in the field.  The ingot was held at the bottom of a vanadium can by a rolled sheet of cadmium.  Magnetic field steps were chosen based on magnetisation measurements.  The incommensurate magnetic structure and its variation with temperature and applied magnetic field are difficult to resolve accurately without a single crystal sample and will not be discussed in detail here.

\section{Results}
\label{sec:Results}
\subsection{Structure and magnetism}
\label{subsec:Magn}
%%%%%%%%%%%%%%%%%%%%%%%%%%%%%%%%%%
%%%%%%%%%%% Neel heat capacity %%%%%%%%%%%%%
%%%%%%%%%%%%%%%%%%%%%%%%%%%%%%%%%%
\begin{figure}
\includegraphics[width=\columnwidth]{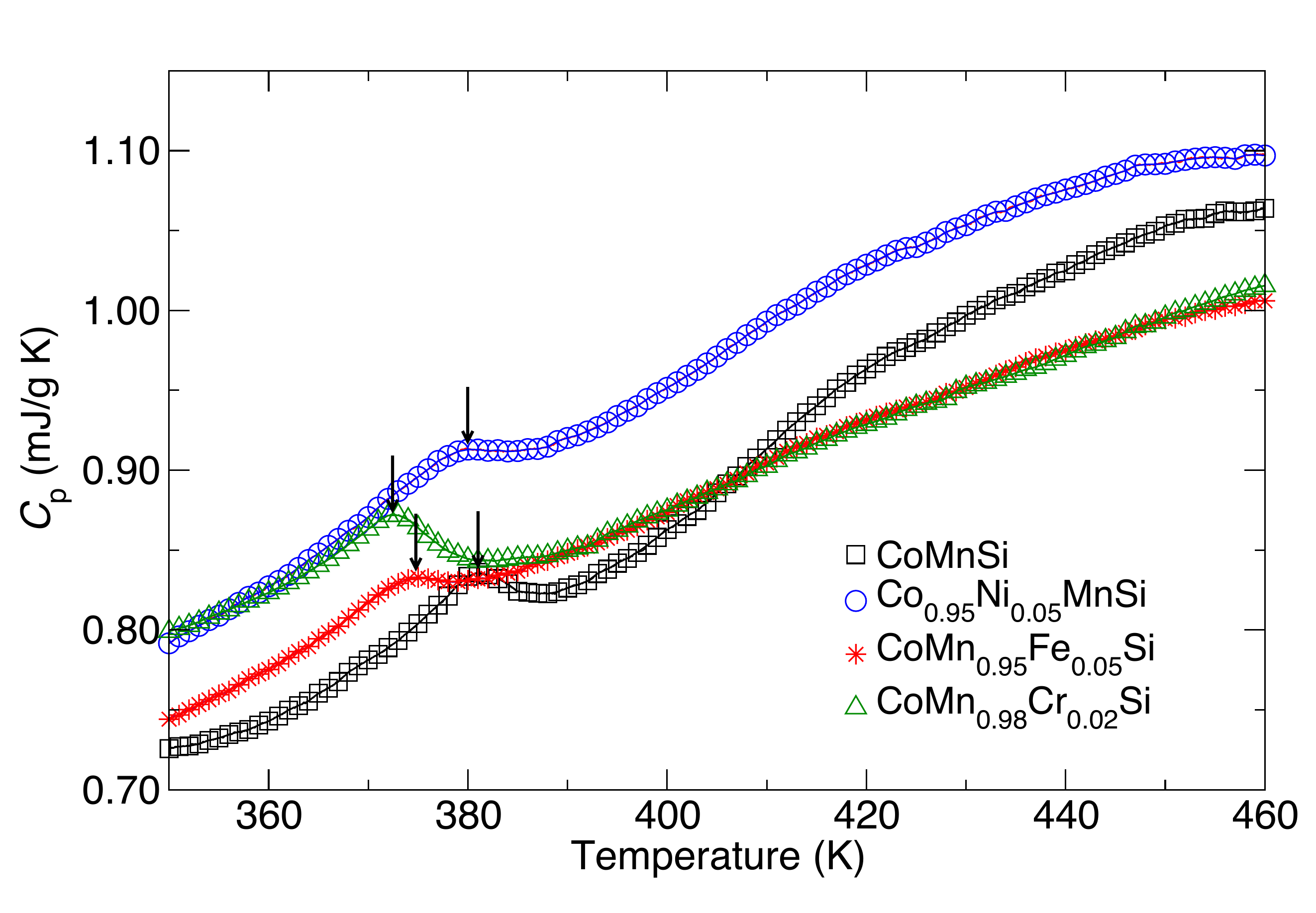}
\caption{(Color online) Specific heat capacity as a function of temperature for four CoMnSi-based alloys. The anomaly around 380~K is assigned to the N\'{e}el phase transition, {\TN}. Measurements were conducted in zero magnetic field.  Despite the small difference in N\'{e}el temperature (shown by the arrows), the alloys have quite different magnetic characteristics and strong variations in thermal expansion.  Symbols and colours are the same as in Figs.~\ref{fig:spont_ms} and \ref{fig:d1d2}.
\label{fig:heatcapacity_lowT}}
\end{figure}
%%%%%%%%%%%%%%%%%%%%%%%%%%%%%%%%%%

\begin{figure}
\includegraphics[width=\columnwidth]{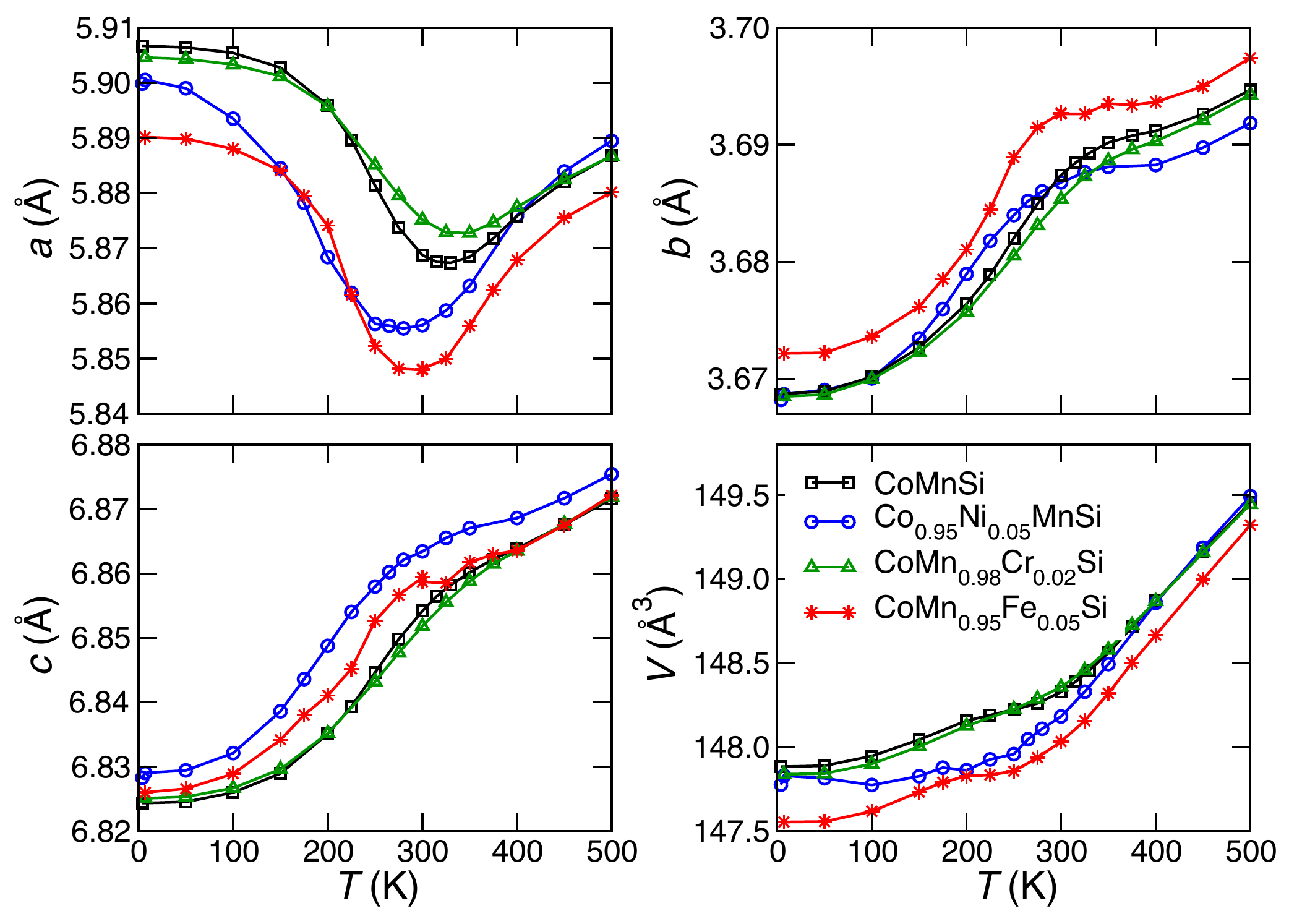}
\caption{(Color online) Temperature dependence of lattice parameters measured on powders using HPRD.  Similar features are seen in all 4 samples.  The large change in cell parameters, including an $a$ axis NTE, occurs well below the N\'{e}el transition temperature, {\TN}.  Such features are strongest in the Ni- and Fe-substituted samples.
\label{fig:spont_ms}}
\end{figure}

The N\'{e}el transition in zero magnetic field is continuous, as evidenced by a broad peak in heat capacity measurements that occurs at {\TN}$\sim$380~K for all samples (Figure~\ref{fig:heatcapacity_lowT}).   Despite the small difference in N\'{e}el temperature between the alloys, there is a strong variation in thermal expansion behaviour with composition.  Figure~\ref{fig:spont_ms} shows the zero-field thermal expansion of {\CMS}, {\CNMS}, {\CMCS} and {\CMFS} from HRPD data.  These are the four main samples involved in this study. As in our previous study~\cite{barcza_2010a}, diffraction data were first recorded at the lowest temperature, before heating the sample to the other temperatures at which data was taken.   If additional patterns were collected the sample was first cooled down to base temperature and then heated up to the desired temperature in order to eliminate any possible effects of thermal hysteresis.   At each step the temperature was equilibrated for 20~min and a total neutron current of either 75~$\mu$A or 60~$\mu$A was collected for each frame.   We used the GSAS code~\cite{larson_1994} to perform a Rietveld refinement of the crystal structure.   Room temperature structural parameters are shown in Table~\ref{StructureTable}.  
%%%%%%%%%%%%%%%%%%%%%%%%%%%%%%%%%%
%%%%%%%%%%% Spont ms figure %%%%%%%%%%%%%%%
%%%%%%%%%%%%%%%%%%%%%%%%%%%%%%%%%%
%%%%%%%%%%%%%%%%%%%%%%%%%%%%%%%%%%

%%%%%%%%%%%%%%%%%%%%%%%%%%%%%%%%%%
%%%%%%%%%%% d1d2  figure %%%%%%%%%%%%%%%
%%%%%%%%%%%%%%%%%%%%%%%%%%%%%%%%%%
\begin{figure}
\includegraphics[width=\columnwidth]{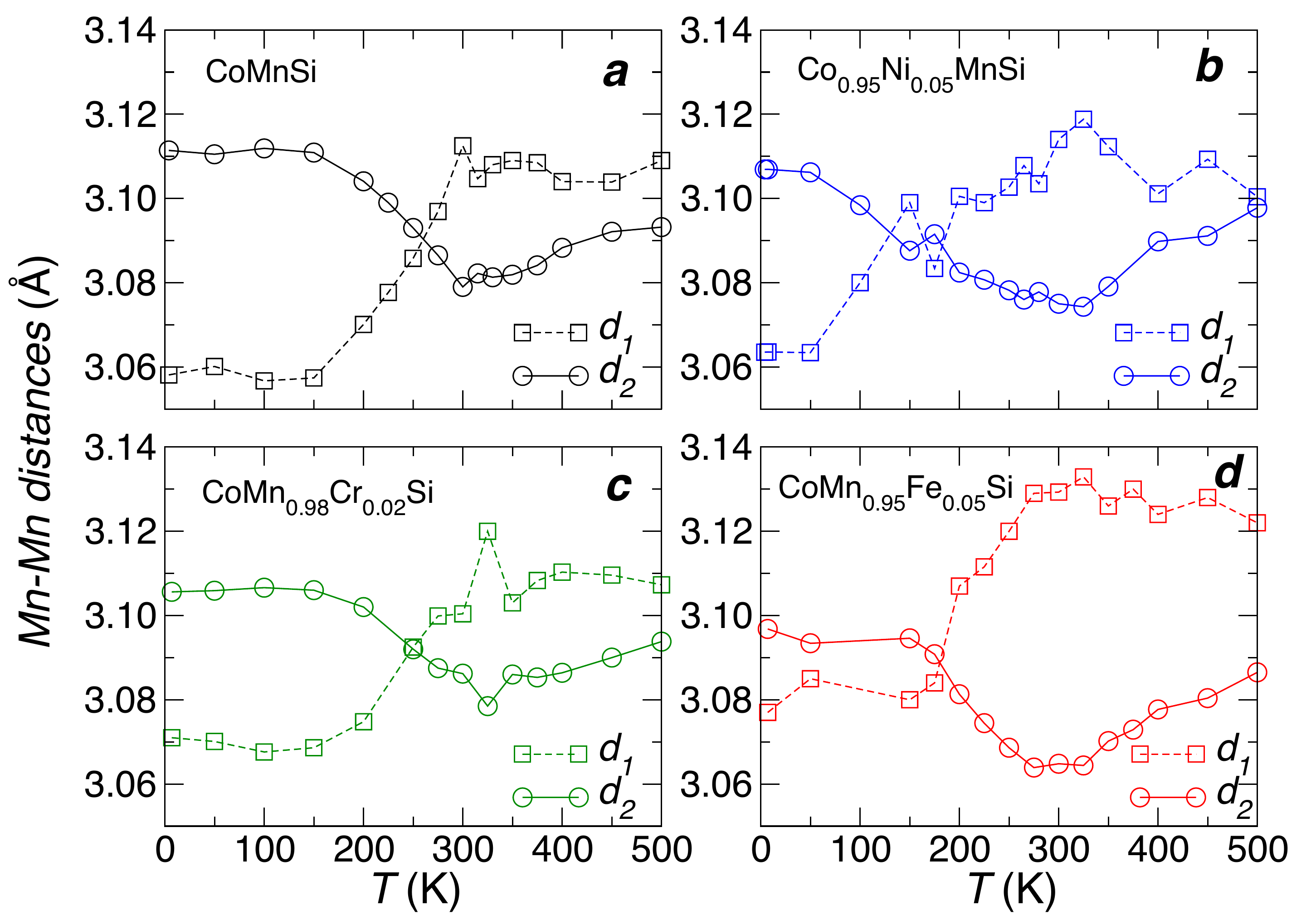}
\caption{(Color online) Manganese nearest-neighbor distances as a function of temperature for the same four samples as in Fig.~\ref{fig:spont_ms}.  Distance {\done} is between different chains of manganese and {\dtwo} is between Mn on the same chain.
\label{fig:d1d2}}
\end{figure}
%%%%%%%%%%%%%%%%%%%%%%%%%%%%%%%%%%

Figure~\ref{fig:d1d2} shows the temperature evolution of the two nearest neighbor Mn-Mn distances, {\done} and {\dtwo} in zero field, obtained from Rietveld refinement of the same data.  {\done} and {\dtwo} are the distances between different chains of manganese atoms, and between Mn atoms on the same chain, respectively.~\cite{barcza_2010a}  We previously identified {\done} as the key structural parameter that determines whether Mn-based orthorhombic magnets in the {\it Pnma} space group have a ferromagnetic ground state or an antiferromagnetic one~\cite{sandeman_2006a, gercsi_2010a}, since, in a certain range of {\done} the coupling between Mn-Mn atoms on different chains is antiferromagnetic.  The temperature variation of {\done} in differently substituted compounds is therefore of interest here.

\begin{table*}
\begin{tabular}{cccccc}
sample ID &  cms38-a &  cms40-a &  cnms39-a &  cmfs33-a &  cmcs41-a
\tabularnewline
\hline
formula & CoMnSi & CoMnSi & {\CNMS} & {\CMFS} & {\CMCS} \tabularnewline
$T_{\rm anneal}$ (K) / $t$ (h) & 1223 / 60 & 1223 / 60 & 1223 / 60 & 1123 / 60 & 1223 / 60        \tabularnewline
2$^{\rm nd}$ phase & t & n& n & y & n \tabularnewline
$a$ ($\AA$) & 5.8688 & 5.8689 & 5.8565 & 5.8480 & 5.8752\tabularnewline
$b$ ($\AA$) & 3.6874 & 3.6916 & 3.6871 & 3.6927 & 3.6854\tabularnewline
$c$ ($\AA$) & 6.8542 & 6.8591 & 6.8639 & 6.8594 & 6.8518 \tabularnewline
$V$ ($\AA^{3}$) & 148.3288 & 148.6063 & 148.2152 & 148.1286 & 148.3577 \tabularnewline
$x_{\rm Co}$ & 0.1558 & 0.1561& 0.1555 & 0.1629 & 0.1550 \tabularnewline
$z_{\rm Co}$ & 0.0605 & 0.0532 & 0.0608 & 0.0611 & 0.0590 \tabularnewline
$x_{\rm Mn}$ & 0.0218 & 0.0209 & 0.0244 & 0.0243 & 0.0221 \tabularnewline
$z_{\rm Mn}$ & 0.6820 & 0.6807 & 0.6816 & 0.6824 & 0.6810 \tabularnewline
$x_{\rm Si}$ & 0.2721 & 0.2712 & 0.2707 & 0.2692 & 0.2715 \tabularnewline
$z_{\rm Si}$ & 0.3733 & 0.3768 & 0.3740 & 0.3729 & 0.3739 \tabularnewline
$R ({\rm wp}) (\%) $ & 7.8 & 4.9 & 6.4 & 6.2 & 5.8 \tabularnewline
$T_{\rm struct}$ &1190~\cite{johnson_1975a} &1190~\cite{johnson_1975a} & & $\sim$1168 & \tabularnewline
\hline
\end{tabular}
\caption{Table of room temperature structural data, derived from Rietveld refinement of HRPD neutron diffraction data (GEM data for cms40-a).  Also shown are: annealing conditions ($T_{\rm anneal}$ and duration, $t$); presence of a 2$^{\rm nd}$ (hexagonal) minority phase (yes/no/trace); goodness of the Rietveld refinement ($R_{\rm RW}$) and the high temperature martensitic phase transition temperature $T_{\rm struct}$ (where known).}
\label{StructureTable}
\end{table*}

\begin{figure}
\includegraphics[width=\columnwidth]{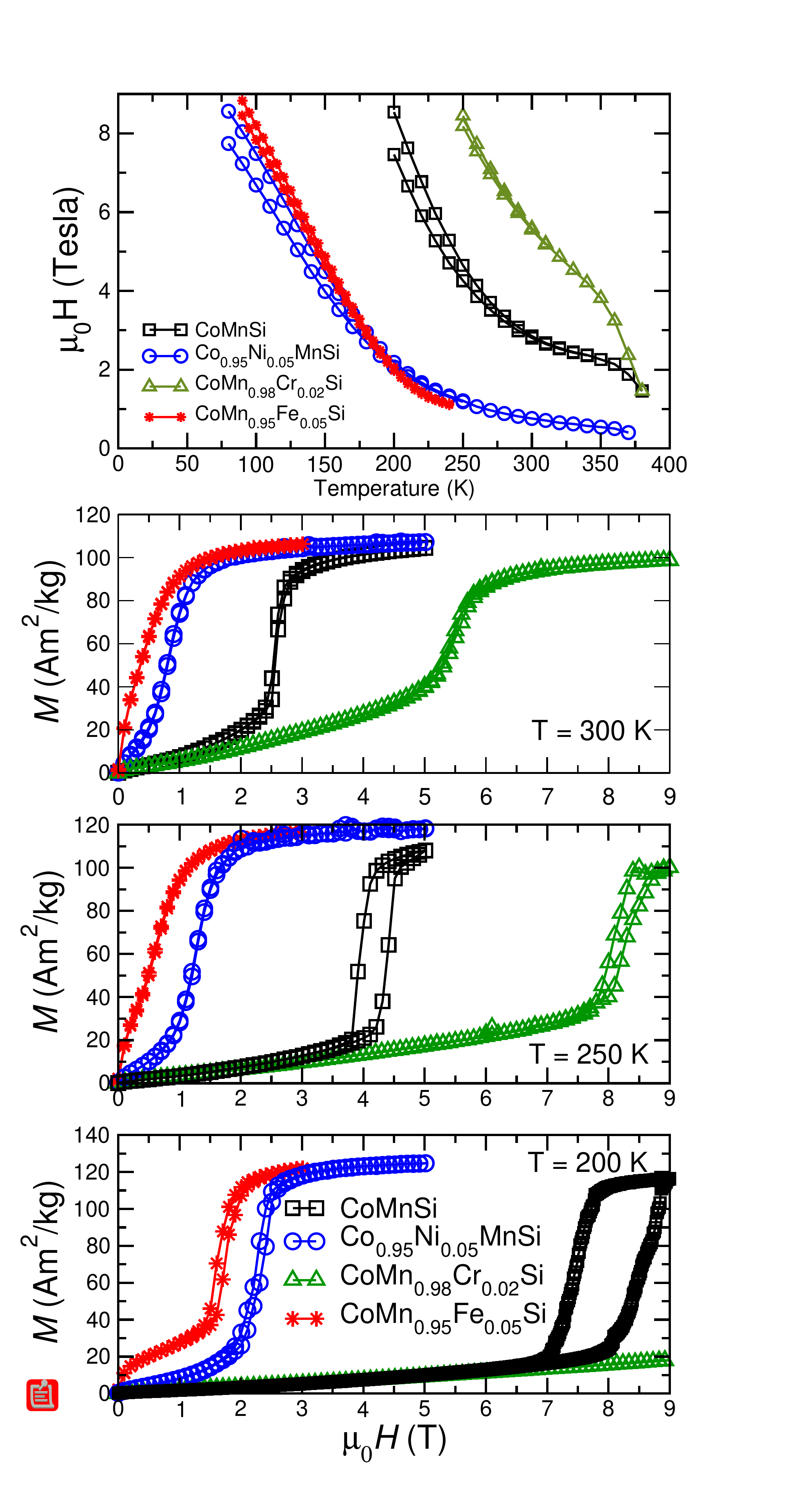}
\caption{(Color online) Upper plot: critical field vs temperature phase diagram for the metamagnetic transition of the 4 main alloys studied here.  Lower 3 plots: isothermal magnetisation data at 300~K, 250~K and 200~K as examples of the data used to construct the phase diagram.  Tricritical points are seen from the narrowing of the hysteresis.  Data for the cms40-a sample of CoMnSi are shown.}
\label{fig:MH4alloys}
\end{figure}

\begin{figure}
\includegraphics[width=\columnwidth]{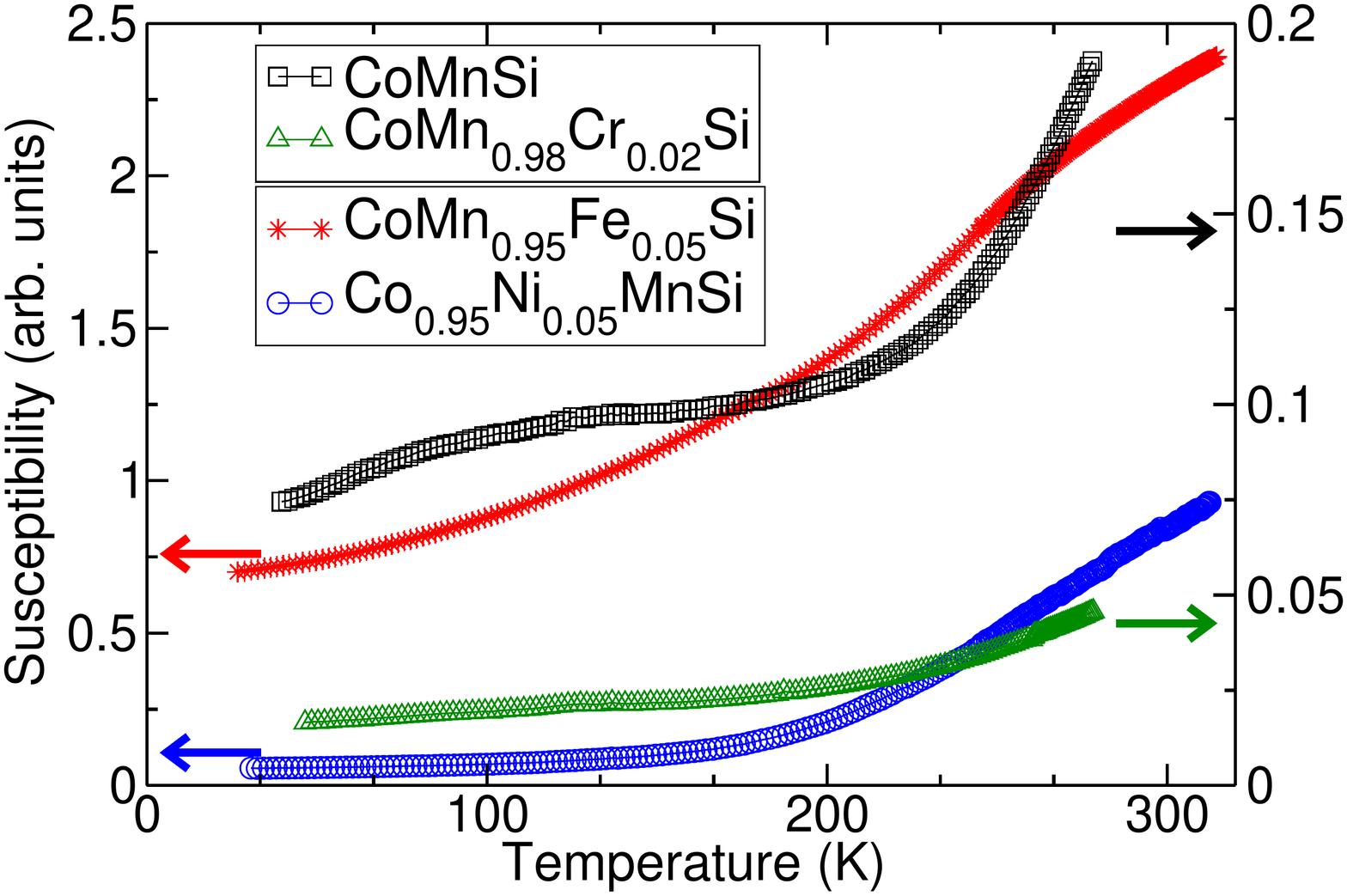}
\caption{(Color online) Magnetic susceptibility of the 4 main alloys studied here measured at 137~Hz in a 100~A/m field.  The susceptibility of CoMnSi and {\CMCS} is lower than the other two alloys, and are plotted on a separate axis.}
\label{fig:mag_susc}
\end{figure}

The structural data for {\CMCS} and {\CMFS} allow us to compare samples in which the helical antiferromagnetism is strengthened or supressed, as can be concluded from the magnetisation data, which we now present.  Figure~\ref{fig:MH4alloys} compares isothermal magnetisation curves at 300, 250 and 200~K in fields of up to 8~Tesla for the same four samples as in Figure~\ref{fig:spont_ms}.    We see that, relative to the stoichiometric material, Cr-substitution increases the critical fields.   Tricritical points are observed in all the alloys, and can be seen from the onset of hysteresis in the magnetisation curves in Fig.~\ref{fig:MH4alloys} and the critical field vs temperature phase diagram that can be constructed from them.  Figure~\ref{fig:mag_susc} shows the temperature variation of magnetic susceptibility measured at 137~Hz in a 100~A/m field.  The Fe- and Ni-substituted alloys have approximately 10 times greater initial susceptibility than CoMnSi, indicating weakened antiferromagnetism (AFM) and greater proximity to ferromagnetism (FM).

%%%%%%%%%%%%%%%%%%%%%%%%%%%%%%%%%%
%%%%%%%%%%% M(H) figure %%%%%%%%%%%%
%%%%%%%%%%%%%%%%%%%%%%%%%%%%%%%%%%
%%%%%%%%%%%%%%%%%%%%%%%%%%%%%%%%%%
%%%%%%%%%%%%%%%%%%%%%%%%%%%%%%%%%%
%%%%%%%%%%% (M,H) and (H,T) figure %%%%%%%%%%%%
%%%%%%%%%%%%%%%%%%%%%%%%%%%%%%%%%%
%%%%%%%%%%%%%%%%%%%%%%%%%%%%%%%%%%
Considering Figs.~\ref{fig:heatcapacity_lowT}, \ref{fig:spont_ms}, and \ref{fig:MH4alloys} together we see that the features of negative expansion along $a$, and positive expansion along $b$ and $c$ are found in all alloys, as well as the giant magneto-elasticity of between 1\% and 2\% change in Mn-Mn separation with temperature, well below {\TN}.  However, the largest changes in $a$ and in {\done} are seen in the Ni- and Fe-substituted materials, which also have the lowest critical fields.    Cr would seem to increase the critical fields and suppress the magneto-elastic interactions that lead to metamagnetism and tricriticality.

\subsection{Entropy change and magnetostriction}
Our interest in tricriticality stems from a desire to understand, control and tune the feedback between structure and magnetism which seem to be key to the giant magneto-elasticity and large magnetocaloric effect found in these materials.  We focus on the temperature range in which the magnetic field induces a metamagnetic transition, below {\TN}, with an associated inverse magnetocaloric effect.  (There is a smaller conventional MCE associated with the depolarisation above {\TN}).   Isothermal magnetisation data presented above are used to callculate isothermal entropy change in each material, in a given field change.  We use the Maxwell relation:

\begin{equation}
\Delta S(T,\Delta H) = \int_{0}^{\Delta H} \biggl( {\partial M \over \partial T} \biggr)_{H} dH \, ,
\label{Eq:Maxwell}
\end{equation}
which can subtlely overestimate the equilibrium phase change entropy around hysteretic phase transitions.~\cite{amaral_2009a}  Such signs of irreversibility are measurable in CoMnSi~\cite{bratko_2011a} but are not the subject of the current study.  Instead we use the Maxwell-derived entropy change to compare materials since thermomagnetic hysteresis is only evident above the tricritical field.

\begin{figure}
\includegraphics[width=\columnwidth]{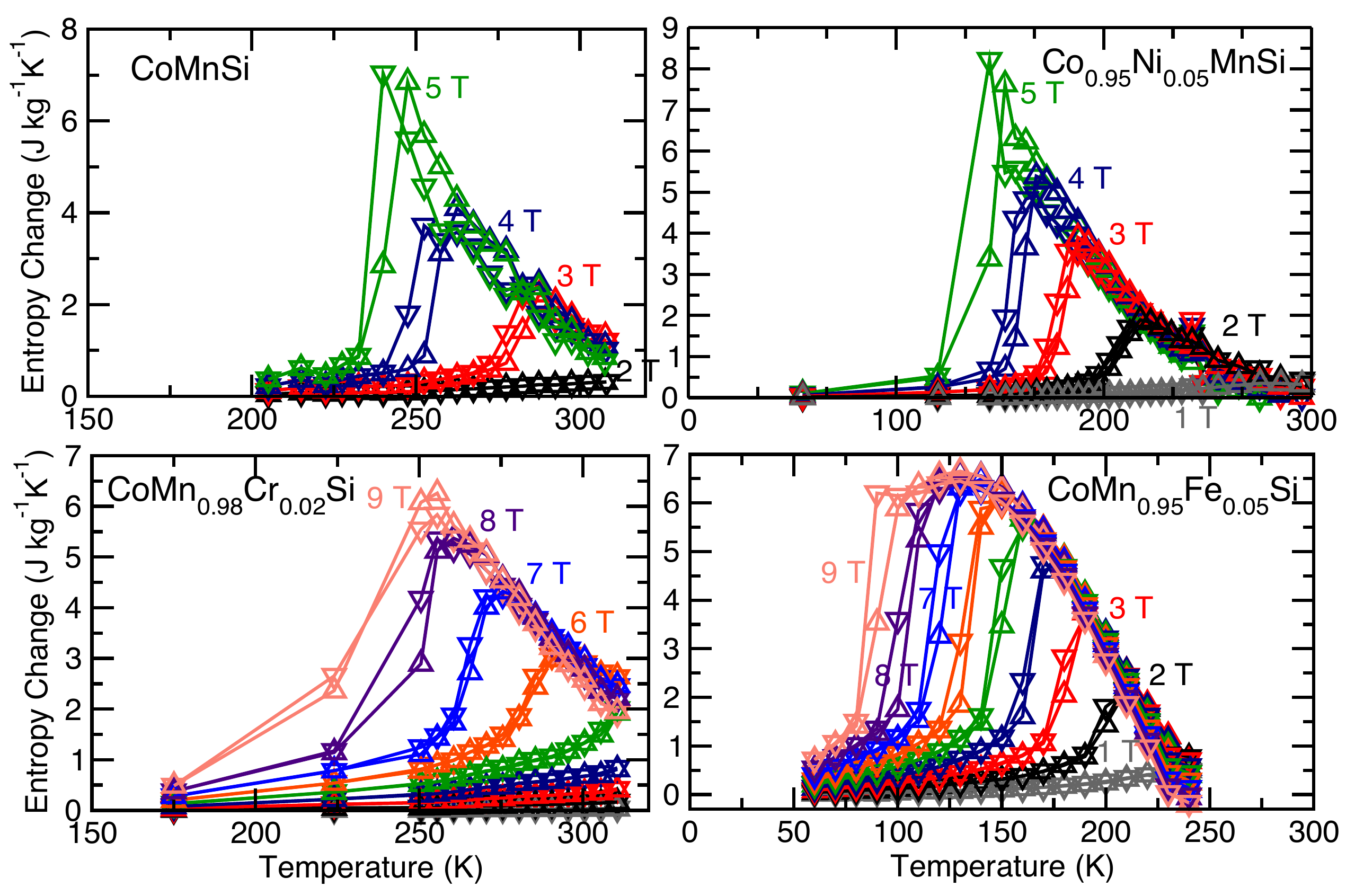}
\caption{(Color online) Isothermal entropy change of the 4 main alloys studied in integer field steps up to 5~Tesla (upper plots) or 9~Tesla (lower plots).  We see that increased proximity to a room temperature tricritical point in Fe- and Ni-doped samples increases the low field entropy change.  The color scheme used is the same in all 4 sub-plots, and up- or down-triangles represent the effect of applying the Maxwell relation to increasing-field or decreasing-field $M(H)$ data.}
\label{fig:DS4alloys}
\end{figure}

Figure~\ref{fig:DS4alloys} shows the derived isothermal entropy changes for all four alloys, in fields of up to 5~Tesla (for CoMnSi and {\CNMS}) or 9~Tesla for the two alloys most magnetically distinct from CoMnSi: {\CMFS} and {\CMCS}.  As can be expected from the $M(H)$ plots, there is a low-field enhancement of the entropy change seen in the Fe-and Ni-substituted materials when compared with CoMnSi.  The Fe-substituted case was presented previously~\cite{morrison_2008a}, with a focus on the entropy change measured directly by calorimetric methods on a 100~$\mu$m-sized fragment.  The additional data taken here enables us to clarify a statement made there.  On increasing the maximum magnetic field strength the metamagnetic phase transition temperature decreases monotonically (see Figure~\ref{fig:MH4alloys}) while the entropy change found at the transition first appears to increase, reaches a maximum, and then becomes smaller.  However, the apparent downturn in the entropy change is an analysis artefact when a hysteretic phase transition is broadened, and hence incomplete, due to effects such as polycrystalline inter-grain strain.  As an example, in a field change of 9~Tesla the lowest temperature at which the phase transition is complete in an increasing field in {\CMFS} is 95~K, which in fact corresponds to the transition temperature for a 8.5~Tesla field.  In addition the magnetization jump and the gradient of the magnetic phase line in ($H,T$) space are both approximately constant from 3~Tesla $<\mu_{0}H <9$~Tesla, so from a Clausius-Clapeyron analysis a significant downturn in entropy change is not expected.
%%%%%%%%%%%%%%%%%%%%%%%%%%%%%%%%%%
%%%%%%%%%%% DS of 4 alloys figure %%%%%%%%%%%%
%%%%%%%%%%%%%%%%%%%%%%%%%%%%%%%%%%
%%%%%%%%%%%%%%%%%%%%%%%%%%%%%%%%%%

Turning to the contributions to the entropy change, we previously concluded that any change in lattice entropy is negative and so adds in opposition to any change in magnetic or electronic entropy.~\cite{morrison_2008a}  However, the signal used to provide a measure of volume change in those experiments was the change in resistance of the calorimeter membrane.  Since the thermal expansion of CoMnSi-based materials is highly aniostropic and the fragment was very small we could not be certain of the size or sign of the volume change in our materials.  Likewise, in previous capacitance dilatometry studies of {\CNMS}~\cite{barcza_2010a}, textured polycrystals were used and so the true dependence of sample volume on magnetic field was unknown.  Therefore, we here examine the magnetostriction of our materials  directly using neutron diffraction and establish the origin of large, competing, changes in entropy that constitute the metamagnetic phase transition.

%%%%%%%%%%%%%%%%%%%%%%%%%%%%%%%%%%
%%%%%%%%%%% MS - CMS  figure %%%%%%%%%%%%%%%
%%%%%%%%%%%%%%%%%%%%%%%%%%%%%%%%%%
\begin{figure}
\includegraphics[width=\columnwidth]{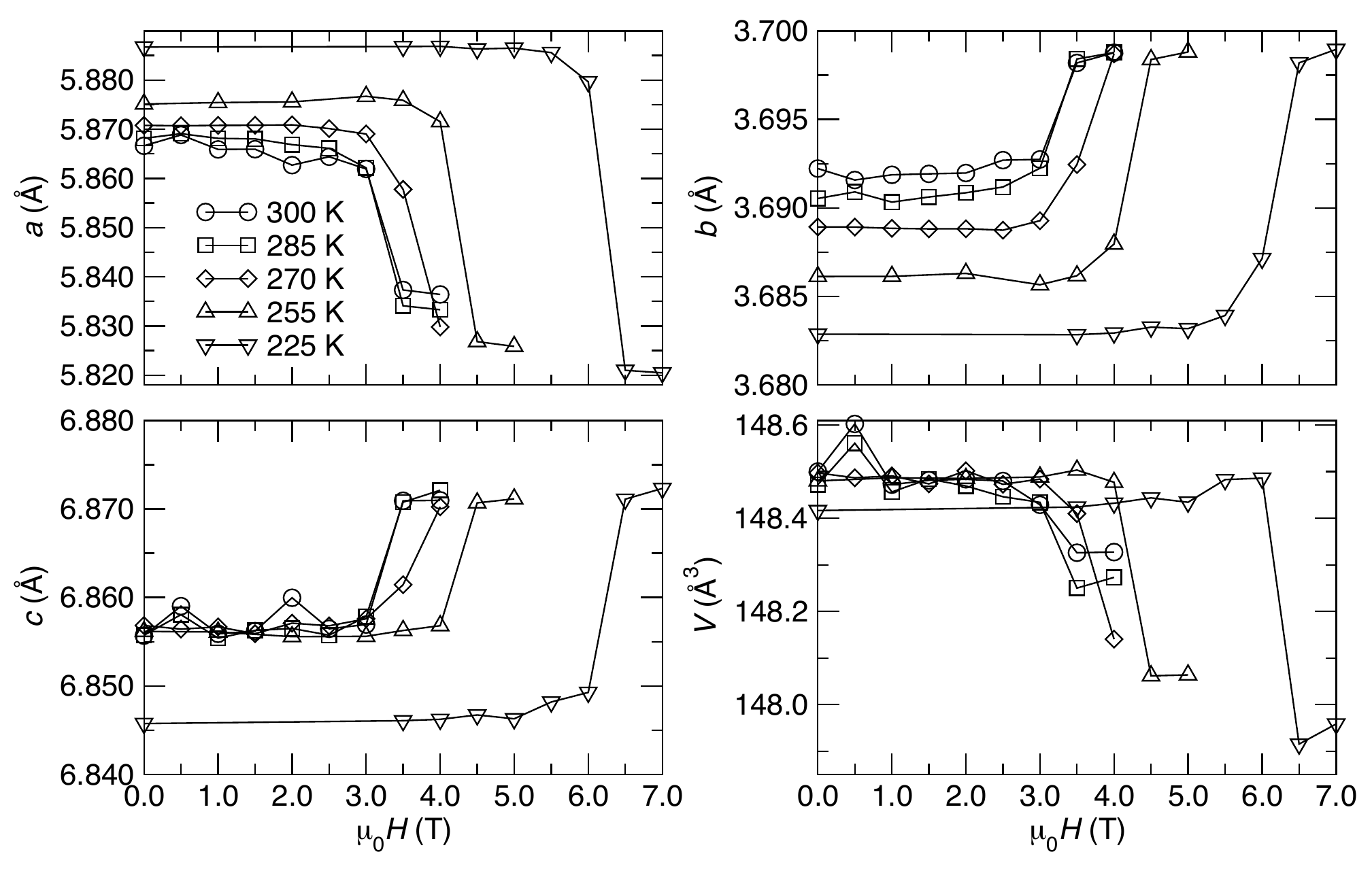}
\caption{Magnetic field dependence of lattice parameters of CoMnSi (cms40-a) as a function of temperature. The zero field values agree well with the data obtained from HRPD on a different sample of CoMnSi (cms38-a). The magnitude of the lattice parameter changes increases with decreasing temperature.}
\label{fig:neutronMS-CMS}
\end{figure}
%%%%%%%%%%%%%%%%%%%%%%%%%%%%%%%%%%

%%%%%%%%%%%%%%%%%%%%%%%%%%%%%%%%%%
%%%%%%%%%%% MS - CNMS  figure %%%%%%%%%%%%%%%
%%%%%%%%%%%%%%%%%%%%%%%%%%%%%%%%%%
%\begin{figure}
%\includegraphics[width=\columnwidth]{cnms39-a_magnetostr_eps.pdf}
%\caption{Forced magnetostriction of {\CNMS} (cnms39-a) from neutron powder diffraction in a magnetic field at two %different temperatures.}
%\label{fig:neutronMS-CNMS}
%`\end{figure}
%%%%%%%%%%%%%%%%%%%%%%%%%%%%%%%%%%

Figure~\ref{fig:neutronMS-CMS} shows the transverse magnetostriction~\cite{ms_comment} in $a$, $b$, $c$, and volume for {\CMS} measured on GEM.   {\CMS} diffraction patterns were collected at 300, 285, 270, 255 and 225~K during cooling with a total counting time of around 30~minutes per pattern.    We see from Fig.~\ref{fig:neutronMS-CMS} that magnetostriction has similar properties to thermal expansion: namely that $a$-axis magnetostriction is negative while that along $b$ and $c$ is positive.  The resulting volume change is small in fields of up to 3~Tesla, above which the $a$-axis behaviour dominates and a large negative magnetostriction is observed.  In the highest applied magnetic fields the lattice parameters assume values never reached during thermal expansion measurements of Figure~\ref{fig:spont_ms}.  For example, in a magnetic field of 6~Tesla the $a$-axis shrinks almost twice as much as it does on changing the temperature from 4~K to 330~K.  On the other hand,  the deformation along the $b$-and $c$-axes induced by the same magnetic field change are comparable to the effect of temperature.  This suggests that the magnetostriction is  influenced predominantly by changes in magnetic order along the a-axis.

The field-induced change in lattice entropy, as discussed in section~\ref{sec:Discussion} below, is thus expected to be large and negative above $\sim$3~Tesla.  The origin of the positive entropy change that dominates the overall MCE is now presented by examining the electronic degrees of freedom via the Sommerfeld coefficient, $\gamma$.  In the metamagnet Fe-Rh two strategies were previously adopted to examine $\gamma$, both using low temperature heat capacity measurements.  The first approach compared the $\gamma$ values of compositions with ferromagnetic (FM) groundstates to those which were antiferromagnets (AFM).~\cite{tu_1969a}  The FMs were found to have significantly higher $\gamma$ values: $\Delta \gamma = \gamma_{FM} - \gamma_{AFM} \sim$6-7~mJ/mol~K$^2$.  A similar value of  $\Delta \gamma \sim14$~mJ/mol~K$^2$ was found for a lower-temperature metamagnet, CeFe$_{0.9}$Co$_{0.1}$.~\cite{wada_1993a}   The second approach examined the magnetic field-induced change in $\gamma$ in metamagnetic compositions of (Fe$_{1-x}$Ni$_{x}$)$_{0.49}$Rh$_{0.51}$ with $x\sim0.03$.~\cite{kreiner_1998a}  Again a similar increase in $\gamma$ was seen on entering the high magnetisation state, thus leading to the conclusion that the giant MCE at the metamagnetic transition was due in large part to an increase in the electronic degrees of freedom, and therefore to a change in the density of states at the Fermi energy.   This increased density of states was a hallmark of the high magnetisation state, brought about either by chemical substitution or by a magnetic field.

%%%%%%%%%%%%%%%%%%%%%%%%%%%%%%%%%%
%%%%%%%%%%% Gamma C(T) Figure %%%%%%%%%%%%
%%%%%%%%%%%%%%%%%%%%%%%%%%%%%%%%%%
\begin{figure}
\includegraphics[width=\columnwidth]{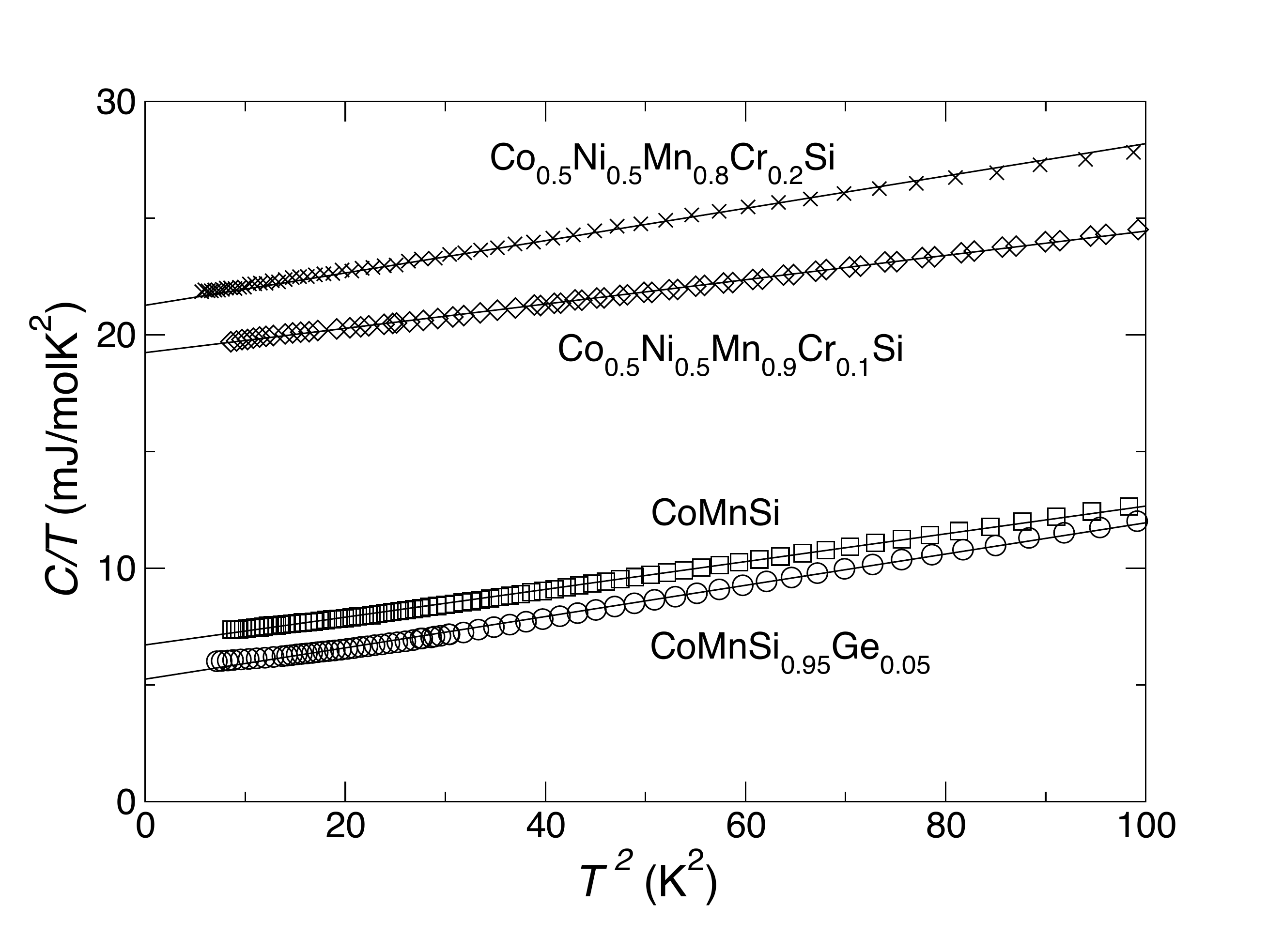}
\caption{Low temperature specific heat data for four CoMnSi-based metamagnets: the $C/T$ axis intercept is the Sommerfeld coefficient, $\gamma$, and is tabulated in Table~\ref{tab:gamma}.}
\label{fig:lowT_Cp}
\end{figure}
%%%%%%%%%%%%%%%%%%%%%%%%%%%%%%%%%%

We employ the first approach here since the fields required to induce metamagnetism at very low temperatures in CoMnSi are much higher than those in Fe-Rh. In Figure~\ref{fig:lowT_Cp} we show specific heat data on four different CoMnSi-based compositions.  Antiferromagnetic compositions are represented by CoMnSi and CoMnSi$_{0.95}$Ge$_{0.05}$; the synthesis and magnetic characterisation of the latter composition was described previously.~\cite{sandeman_2006a}  In order to find a CoMnSi-based material with a ferromagnetic groundstate, a higher level of chemical substitution than is present in any of the samples presented above is required.  Therefore two new compositions were prepared for this experiment using the same protocol as given in Section~\ref{sec:Experimental}.  These are Co$_{0.5}$Ni$_{0.5}$Mn$_{0.8}$Cr$_{0.2}$Si and Co$_{0.5}$Ni$_{0.5}$Mn$_{0.9}$Cr$_{0.1}$Si.  They were found to have high magnetisation groundstates (not shown here) which are assumed to be ferromagnetic.

We can see from Figure~\ref{fig:lowT_Cp} that, as with Fe-Rh, there is a very large increase in $\gamma$ when comparing a FM composition with a AFM one.  Compared to Fe-Rh, the change in $\gamma$ in CoMnSi-based materials would seem to be much larger, at around 15~mJ/mol K$^2$.  

\section{Discussion}
\label{sec:Discussion}
\subsection{Giant magneto-elasticity and tricriticality}
\label{sec:tricriticality}
The results presented here enable us to establish and quantify several properties of CoMnSi-based metamagnets for the first time.  From section~\ref{subsec:Magn} we see that the effects of Fe or Ni doping are opposite to that of Cr.    Critical metamagnetic fields are reduced be either of the former, and enhanced be the latter.  The peak in heat capacity at {\TN} is suppressed by Fe or Ni, and enhanced by Cr.   The change in lattice parameters and in Mn-Mn distances (in particular {\done}) become stronger with Fe, are shifted down in temperature by Ni, but are suppressed by Cr.  If we ascribe the strength of heat capacity jump, size of metamagnetic critical field and decreased magneto-elastic coupling to the presence of antiferromagnetism with increased anisotropy then, as we now outline, the above data can be viewed as self-consistent.

In CoMnSi, the incommensurate helical magnetic ordering wavevector is (0 0 $q$), where $q\sim 0.4$ at low temperature.  One possibility is that Fe or Ni doping reduce the value of $q$, bringing the material closer to ferromagnetism and reducing the metamagnetic critical field.   Cr would seem to have the opposite effect of stiffening the helical arrangement of moments, presumably as a result of strengthening the antiferromagnetic exchange that is mediated by the inter-chain Mn-Mn separation {\done}.   We previously proposed a simple model of magnetic groundstates in Mn-based {\it Pnma} alloys based on MnP, where low values of {\done} suppress any magnetic order, and then, on increasing {\done}, two boundaries between different magnetic groundstates are seen.  First FM, then AFM and finally FM states are stabilised as {\done} increases.~\cite{gercsi_2011a}  The {\done} values at the theoretical FM/AFM phase boundaries~\cite{gercsi_2011a} are $\sim$3~\AA~and $\sim$3.36~\AA.  The compounds presented in Figs.~\ref{fig:spont_ms} and \ref{fig:d1d2} have {\done} values between 3.06 and 3.08~\AA~at 4~K and are therefore on the lower boundary between low volume ferromagnetism and higher volume antiferromagnetism.  Disordered local moment first-principles calculations have recently confirmed that an increase in {\done} raises the metamagnetic critical field~\cite{staunton_2012a}.  {\CMCS} has a higher value of {\done} at low temperature than either CoMnSi or {\CNMS} and can therefore be thought of as sitting further from the FM/AFM magnetic phase boundary.  The findings of high critical fields and reduced magneto-elastic coupling in this material are therefore self-consistent.  The fact that {\CMFS} has a high value of {\done} at low temperature but has reduced critical fields highlights the importance of changing the  electron count associated with the Mn site, and the relevance of examining the difference in electronic densities of statements, as performed in the next Section.

We therefore conclude that Cr reduces the magneto-elastic coupling in zero field, if indeed that coupling arises from a difference between the value of {\done} in the antiferromagnetic groundstate and in the high magnetisation state and that the reverse is true in the case of the Fe and Ni substitutions studied.    Another interesting possibility is that the Mn moments, which are believed to lie in the ($a$,$b$) plane in CoMnSi, are tilted out of plane by Mn/Fe or Co/Ni substitution to form a canted ferromagnetic structure, thus enhancing the magnetic susceptibility and reducing the (tri)critical fields.  Further neutron diffraction measurements on single crystalline samples are required to examine the relationship between magnetic and crystal structure.  A previous study of the magnetoresistance of CoMnSi has suggested the presence of a spin reorientation transition below the N\'{e}el temperature on the basis of a cusp in the ac susceptibility at $\sim$150~K.~\cite{zhang_2008a}  However, our magnetic susceptibility measurements (Figure~\ref{fig:mag_susc}) show no sign of such a feature at this temperature.

\subsection{Electronic contributions to $\Delta S$}
\label{sec:MCE_elec}
We now present an analysis of the components of the isothermal entropy change, by using as an example the transition in CoMnSi at a temperature of 240~K ($\sim$5~Tesla), where the change of volume is around $-0.3$\,\% (Figure~\ref{fig:neutronMS-CMS}) and $\Delta S$ is around 7\,J\,kg$^{-1}$\,K$^{-1}$.  In order to deconstruct the entropy change at this transition we start with the electronic contribution and then proceed to the role of the lattice.   It is not straightforward to calculate magnetic entropy contributions without detailed knowledge about the magnetic structure in both magnetic states and whether there is a jump in the site moment as found in other systems such as FeRh and (Mn,Fe)$_{2}$(P,Z).~\cite{shirane_1963a,shirane_1964a,dung_2011a}  Without any evidence for such changes in site moment, magnetic contributions to the phase transition between two ordered states are expected to be small and will not be considered other than through a change in the density of electronic states (DOS).  Hereafter we refer simply to the ``electronic" entropy change, while recognising that DOS changes will contribute to the itinerant part of the magnetic entropy change.

The large change of electronic entropy at the metamagnetic transition is consistent with our earlier theoretical result that a helical magnetic state has a lower density of electronic states at the Fermi energy than a (fictitious) ferromagnetic ground state.~\cite{barcza_2010a}   It is also perhaps consistent with the extraordinary sensitivity of CoMnSi to external parameters such as synthesis conditions and external parameters such as pressure and magnetic field.  We previously showed a variation of 200~K in the low field metamagnetic transition temperature  from literature data.~\cite{sandeman_2006a}  Even in the work presented here, two ingots of the same nominal composition (cms38-a and cms40-a) have slightly different critical fields and temperatures.

A list of the low temperature Sommerfeld coefficients, $\gamma$-coefficients extracted from the data in Figure~\ref{fig:heatcapacity_lowT} is given in Table~\ref{tab:gamma}.  It shows that $\gamma$ is approximately tripled when moving from an antiferromagnetic alloy to a ferromagnetic one. The change in $\gamma$ for a selection of materials that exhibit AFM and FM ground states or field-induced high magnetisation states is also shown.  We note that the scale of $\gamma$ measured in either mJmol$^{-1}$K$^{-2}$ varies greatly between systems and that measurement in mJ$\,$kg$^{-1}$\,K$^{-2}$ reduces this difference in scale.  The magnitude of $\Delta\gamma=\gamma_{FM}-\gamma_{AFM}$ is largest in CoMnSi, when measured in mJkg$^{-1}$K$^{-2}$, highlighting the size and importance of the change electronic density of states in this compound.  Division by number of atoms per mole perhaps equally instructive, yielding $\Delta\gamma\sim$ 4.7, 3.5, 4.4 and 2.1 mJmol$^{-1}$atm$^{-1}$K$^{-2}$ for CoMnSi, Fe-Rh, Ce(Fe,Co)$_{2}$ and (Nd,La)Fe$_{11.5}$Al$_{1.5}$ respectively.

\begin{table}
\begin{tabular}{lcc}
 & $\gamma$  & $\gamma$ \tabularnewline
compound & (mJ\,mol$^{-1}$\,K$^{-2}$) & (mJ\,kg$^{-1}$\,K$^{-2}$) \tabularnewline
\hline
\hline
CoMnSi (AFM) & 6.7 & 46.5 \tabularnewline
CoMnSi$_{0.95}$Ge$_{0.05}$ (AFM) & 5.2 & 36.6\tabularnewline
Co$_{0.5}$Ni$_{0.5}$Mn$_{0.9}$Cr$_{0.1}$Si (FM) & 19.2 & 135.6\tabularnewline
Co$_{0.5}$Ni$_{0.5}$Mn$_{0.8}$Cr$_{0.2}$Si (FM) & 21.3 & 150.8\tabularnewline
\hline
Fe$_{49}$Rh$_{51}$ (AFM)\cite{tu_1969a} & 2.5 & 16\tabularnewline
Fe$_{51}$Rh$_{49}$ (FM)\cite{tu_1969a} & 9.5 & 60\tabularnewline
\hline
Ce(Fe$_{0.9}$Co$_{0.1}$)$_{2}$ (AFM)\cite{wada_1993a} & 36.6 & 145\tabularnewline
Ce(Fe$_{0.9}$Co$_{0.1}$)$_{2}$ (FM, est.)\cite{wada_1993a} & 50 & 198\tabularnewline
\hline
LaFe$_{11.5}$Al$_{1.5}$ (AFM)\cite{fang_2008a}& 193 & 235.6\tabularnewline
Nd$_{0.2}$La$_{0.8}$Fe$_{11.5}$Al$_{1.5}$ (FM)\cite{fang_2008a} & 221.6 & 270.4\tabularnewline
\hline
\hline
\end{tabular}
\caption{Sommerfeld $\gamma$-coefficients for several CoMnSi alloys, compared with literature values found in a range of AFM/FM metamagnets.}
\label{tab:gamma}
\end{table}

It can reasonably be assumed that the electronic heat capacity at the field-induced metamagnetic phase transition in CoMnSi changes by a similar order of magnitude to that found on comparing an antiferromagnetic alloy with a ferromagnetic one.  However, this assumption is oversimplified for two reasons, the impact of which need to be examined briefly. Firstly, the substitution used to generate ferromagnetism is not isoelectronic. This makes a direct comparison of the $\gamma$-coefficients of two compounds nontrivial. Secondly, even if such an experiment can be conducted with isoelectronic samples it can still not be excluded that some change of $\gamma$ comes from volume changes and their impact on the density of states at the Fermi energy. However, this effect scales as the fractional volume change~\cite{jia_2006a}: 
$\Delta S_{\mathrm{el, volume}}=\frac{2}{3}\frac{\Delta V}{V}\gamma T$ and is therefore small ($\sim$~0.018\,J\,kg$^{-1}$\,K$^{-1}$ at 240~K).
Therefore the threefold increase in $\gamma$ observed is unlikely to originate from substitution only, or from volume changes.

A band structure feature is therefore most likely to contribute to such a large difference between the entropy of   the AFM state and field-induced high magnetisation state.  This idea is corroborated by the large difference between the DOS of a non-collinear AFM and that of a collinear FM structure in CoMnSi that we previously calculated from DFT.~\cite{barcza_2010a}    However, the shape of the DOS is crucial here.  It might seem simplest to examine the  average difference between the electronic heat capacity in the AFM and FM alloys which is $\sim$100~mJ\,kg$^{-1}$\,K$^{-2}$ and then make an estimate of the electronic entropy change, as is often done in other materials~\cite{wada_1993a,kreiner_1998a} using $\Delta S_{\mathrm{el}}= T_{\mathrm{t}} \Delta\gamma$.  In the current case, this yields a value of ~24~J\,kg$^{-1}$\,K$^{-1}$ at 240~K which would be an upper limit for $\Delta S_{\mathrm{el}}$ at this temperature.  However, the form of $\Delta S_{\mathrm{el}} \propto T_{\mathrm{t}}$ predicts that the electronic contribution should grow linearly in magnitude with $T_t$ (decreasing field) which contradicts the opposite trend in $\Delta S(T_{\mathrm{t}})$ seen in Figure~\ref{fig:DS4alloys} for most of the field range studied.   In other words, the effects of finite temperature on energy-dependent bandstructure need to be taken into account.

\begin{figure}
\includegraphics[width=\columnwidth]{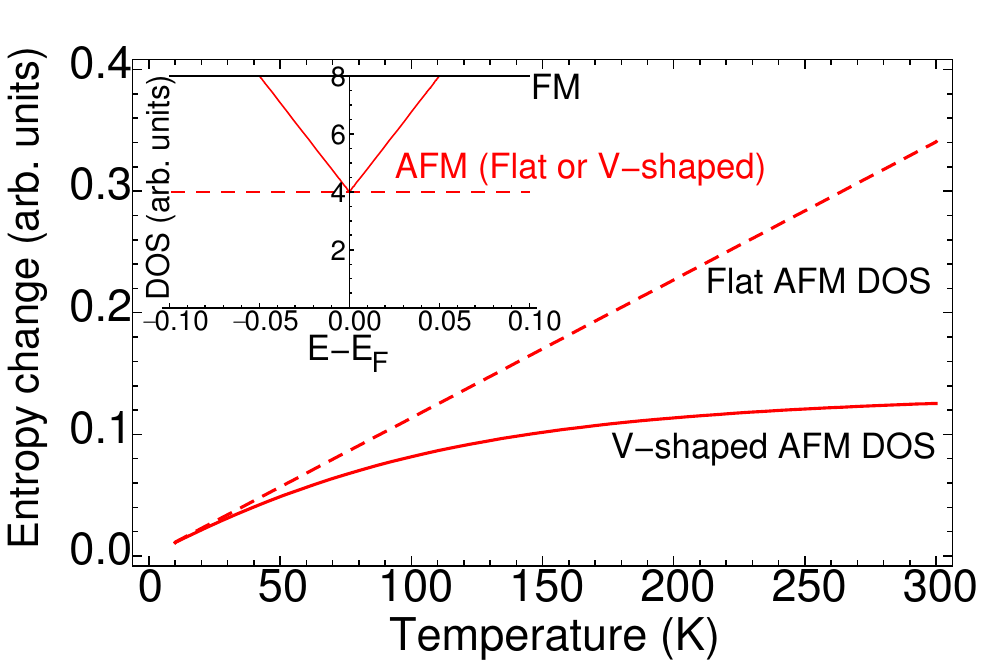}
\caption{A simple V-shaped model DOS for the AFM state near $E_F$ (red solid line, inset) yields a reduced entropy change (red solid line, main figure) relative to that of two energy-independent DOS in both the AFM and FM states).}
\label{fig:DOS-model}
\end{figure}

We can see from our earlier DFT calculations~\cite{barcza_2010a} (Figure~4 of that study) that the AFM DOS has a dip near the Fermi Energy, $E_F$ which has an energy width of order $k_{B}T$ and that the lowest DOS value is only to be found close to $E_F$ itself.   Thermal occupation of the DOS is indeed therefore relevant to the measured electronic entropy change.   We may hence use a simple model of a V-shaped DOS, of width  0.1~eV  and employ the statistical description of fermionic entropy to the FM and the AFM DOS:
\begin{equation}
S_{\mathrm{el}}=-k_{B}\int{dE [f\log f + (1-f)\log(1-f)]}.
\end{equation}
The (dis)occupation of states with DOS higher than the dip at the Fermi energy reduces the entropy difference between FM and AFM states, relative to the simplified model stated above by a factor of two or more in the room temperature range (see Figure~\ref{fig:DOS-model}).  The precise reduction depends on the gradient of the DOS with energy on either side of $E_{F}$.  Parameters here are chosen to match the form of the DOS calculated previously and the approximate increase in DOS at the Fermi level obtained from that calculation and from Table~\ref{tab:gamma}.
%%%%%%%%%%%%%%%%%%%%%%%%%%%%%%%%%%
%%%%%%%%%%% Model of DOS-DS   %%%%%%%%%%%%
%%%%%%%%%%%%%%%%%%%%%%%%%%%%%%%%%%
%%%%%%%%%%%%%%%%%%%%%%%%%%%%%%%%%%

Hence, rather than 24~J\,kg$^{-1}$\,K$^{-1}$ at 240~K, a value closer to 10 J\,kg$^{-1}$\,K$^{-1}$ electronic entropy change would seem reasonable at this temperature.  A fuller examination of the entropy change across a range of temperatures --- including at low fields where the transition to a high moment state is incomplete --- would perhaps require more than the V-shaped AFM-DOS model.  The motivation here is to stress the importance of the DOS profile near the Fermi Energy.  Despite the  appearance of a very large change in $\gamma$ coefficient, the accessible change in electronic entropy is limited by the varation of AFM DOS near to $E_F$.

In order to confirm the large electronic entropy change in CoMnSi further experiments on isoelectronically substituted compounds are encouraged. A possible alloy series could be germanium rich CoMnSi$_{1-x}$Ge$_{x}$ which are ferromagnetic for large \emph{x}.~\cite{niziol_1980a} By changing the germanium content it may be possible to tailor the magnetic properties of the alloy in order to access the transition in the magnetic field of a laboratory magnet.

\subsection{Lattice contributions to $\Delta S$}
\label{sub:MCE_lattice}
The lattice contributions to $\Delta S$ can be estimated using the protocol described by Jia et al.~\cite{jia_2006a}  Accordingly, there are two components to the lattice entropy: (i) a phonon contribution due to the shift in Debye temperature caused by the field and (ii) an elastic contribution due to the deformation of the bulk material.  The Debye temperature, $T_{\mathrm{D}}$ can be estimated from a fit of the experimental heat capacity data to the a Debye model for $C_{V}(T)$.  This assumes that $C_{P}(T)$ data taken under isobaric conditions differ only slightly from the isochoric data required for the Debye model.   The difference between isochoric and isobaric $C(T)$ requires knowledge of the isothermal compressibility $\kappa_{\mathrm{T}}$ and the coefficient of linear thermal expansion, $\alpha_{\mathrm{T}}$, of the material: 
\begin{equation}
C_{\mathrm{P}}-C_{\mathrm{V}}=\frac{Tv\alpha_{\mathrm{T}}^{2}}{\kappa_{\mathrm{T}}}
\end{equation}
where $v$ is the molar volume. Due to the small thermal expansion of CoMnSi at low temperatures, this correction does not contribute significantly to the quality of the fit, especially since the model is a rather simple one.  

A Debye temperature in the antiferromagnetic state of $T_{\mathrm{D}}=410$\,K results in the best fit of experimental data to the Debye model for temperatures $T<200$~K.  Strictly speaking the Debye model is only valid for monatomic isotropic materials. In CoMnSi the lattice expands anisotropically which might explain the deviation from the simple model at temperatures above this range. Interestingly the temperature range above which the agreement of experimental heat capacity data with the model starts to get worse coincides with the start of negative thermal expansion along the \emph{a}-axis (Figure~\ref{fig:spont_ms}). 

%where $C_{\mathrm{V}}$ is the heat capacity at constant volume, $R$ is the gas constant, $k_{\mathrm{B}}$ is Boltzmann's constant, $N_{\mathrm{A}}$ is Avogadro's number and $T_{\mathrm{D}}$ the material specific Debye temperature, defined in equation~\ref{eq:Debye_freq}, $\omega_{\mathrm{D}}$ the maximum phonon frequency of the system, and $k_{\mathrm{B}}$ the Boltzmann constant.  The Debye temperature which is needed to calculate $\Delta S_{\mathrm{lat}}$ can  be obtained from a fit of the experimental heat capacity data to the model in equation~\ref{eq:Debye}.  This assumes that $C(T)$ data taken under isobaric conditions differ only slightly from the isochoric data required for Eq.~\ref{eq:Debye}. 
%
%\begin{figure}
%\includegraphics[width=\columnwidth]{Debye_CMS0T}
%\caption[Fit of the heat capacity of CoMnSi to the Debye model.]{A fit of the experimental heat capacity data of CoMnSi (open circles) to the Debye model (line).
%\label{fig:Debye_CMS}}
%\end{figure}

The lattice entropy for a system is then given by~\cite{oliveira_2004a,tari_2003a}:
\begin{eqnarray}
S_{\mathrm{lat}}(T)=-3N\, R\, ln(1-e^{-\frac{T_{\mathrm{D}}}{T}}) \\
+12N\, R\left(\frac{T}{T_{\mathrm{D}}}\right)^{3}\int_{0}^{T_{\mathrm{D}}/T}\frac{x^{3}}{e^{x}-1}dx\
\end{eqnarray}
where $N$ is the number of atoms per mole and $R$ the gas constant. A volume change influences the phonon system and hence the Debye temperature $T_{\rm D}$ by~\cite{taylor_1998a}:

\begin{equation}
\frac{\Delta T_{\mathrm{D}}}{T_{\mathrm{D}}}=-\eta\frac{\Delta V}{V}
\label{eq:gruneisen}
\end{equation}
where $\eta$ is the Gr\"{u}neisen parameter. This parameter depends on the material and is between unity and 3 for many systems. In LaFe$_{13-x}$Si$_{x}$ systems $\eta$ has elsewhere been set to 6 which resulted in the best fit of experimental data to a model similar to the one discussed here.~\cite{jia_2006a} The Gr\"{u}neisen parameter for CoMnSi is unknown. Setting $\eta=3$ as an example yields a change in $T_{\mathrm{D}}$ of about 3.7\,K or a $T_{\mathrm{D}}^{\mathrm{high M}}=413.7\,\mathrm{K}$ in the spin-aligned state.  The lattice entropy change can then be calculated from the difference between $S_{\mathrm{lat}}$ values at a single temperature (240~K) using two different Debye temperatures.  For $\eta=3$ we obtain $\Delta S_{\mathrm{lat}}\simeq-4$\,Jkg$^{-1}$K$^{-1}$. The sign of the lattice entropy change is negative since a higher Debye temperature in the spin-aligned state means that fewer phonon modes are excited at a fixed temperature. 

The elastic contribution, $\Delta S_{\mathrm{ela}}$ examines the deformation of the crystal as a whole. According to Hooke's law the energy difference between two states assuming that the bulk modulus is constant is given by:
\begin{equation}
\Delta U_{\mathrm{ela}}=\frac{1}{2}B\frac{(\Delta V)}{V_{0}}^{2}\label{eq:U_ela}
\end{equation}
where $V_{0}$ is the reference volume.  For CoMnSi at 240~K the volume difference between the AFM and FM-like unit cells is $\Delta V\sim0.5~$\AA$^{3}$.  The unit cell volume in the AFM reference state at 240\,K is 148.34~\AA$^{3}$.  The bulk modulus of CoMnSi is unknown but we might assume that it is equal to that of MnSi, $B=1.63\times10^{7}\,\mathrm{N/cm^{2}}$.~\cite{petrova_2009a} From these values we can obtain a tiny effective latent heat at 240~K of $\Delta S_{\mathrm{ela}}=0.61\,\mathrm{J\, kg^{-1}\, K^{-1}}$.  It should be emphasised that this value is based on an estimated, field- and temperature-independent bulk modulus.

In summary it seems that in CoMnSi the positive field-induced isothermal entropy change is mainly due to an  contribution from the electronic density of states (similar to FeRh-based alloys which also exhibit an antiferro- to ferromagnetic phase transition and estimated here to be 10~J\,kg$^{-1}$\,K$^{-1}$ at 240~K) and counteracting phonon density contributions (e.g. around -4~Jkg$^{-1}$K$^{-1}$ at 240~K).   A naive isotropic analysis of the phonon entropy contribution is clearly not sufficient.  Direct phonon spectroscopy is needed --- as it is for several magnetocaloric systems --- in order to more accurately determine the role of phonons in contributing to the total entropy change determined from magnetic or calorimetric measurements.  The role of the lattice can be extremely important in other materials, either in adding to the magnetic entropy change (Gd-Si-Ge~\cite{pecharsky_2003a} or in counteracting it, as seen here and in La-Fe-Si.~\cite{jia_2006a}  Recent theoretical work by Basso has extended the model of first order phase transitions by Bean and Rodbell to calculate the role of the lattice in the isothermal entropy change at a magnetic field-driven phase transition.~\cite{basso_2011a}  Further extension of such a model to anisotropic systems such as found in (Mn,Fe)$_{2}$(P,Z) and CoMnSi will be of interest in order to relate anisotropy in thermal expansion to magneto-elastic entropy.

\section{Conclusions}
We have found that  different elements have contrasting effects on the position of the tricritical point in ($T,H$) space and that this can be related to the size of the magneto-elastic coupling as manifested by the temperature dependence of manganese nearest-neighbor distances.  The entropy change associated with metamagnetism in CoMnSi is dominated by a very significant change in electronic entropy, compensated by an opposite phonon contribution.  Our work motivates further attempts at single crystal synthesis in order to undertake neutron diffraction experiments to examine the temperature and field evolution of the magnetic structure, as well as neutron spectroscopy to examine the nature of the changing phonon density of states.  The variation in tricritical fields are also  an interesting test bed for finite temperature theories of magnetism.~\cite{hughes_2007a,staunton_2012a}

\begin{acknowledgments}
We thank M. Avdeev, R.Bali, N. Shannon and J.B. Staunton for useful discussions, K. Roberts for help with sample preparation, J. Aryaman for assistance with data analysis and G.G. Lonzarich for the use of sample synthesis facilities.  A.B. would like to thank the EPSRC, The Leverhulme Trust and Camfridge Ltd. for financial support.  K.G.S acknowledges financial support from The Royal Society and thanks Monash University for hosting a research visit.  The research leading to these results has received funding from the European Community's 7th Framework Programme under grant agreement No. 214864 (``SSEEC").
\end{acknowledgments}


\begin{thebibliography}{}
\bibitem{ms_comment} The magnetic field was perpendicular to the scattering plane for the diffraction experiment.
\end{thebibliography}
\end{document}